\documentclass{aastex631}

\usepackage{fancybox} % for draft

\received{December 13, 2022}
\revised{February 1, 2023}
\accepted{February 22, 2023}

\graphicspath{{./}{figures/}}

\begin{document}

\title{Solar Energetic Particle Events with Short and Long Onset Times}

\author{Kosuke Kihara}
\affiliation{Astronomical Observatory, Kyoto University, Sakyo, Kyoto 606-8502, Japan}

\author{Ayumi Asai}
\affiliation{Astronomical Observatory, Kyoto University, Sakyo, Kyoto 606-8502, Japan}

\author{Seiji Yashiro}
\affiliation{Department of Physics, The Catholic University of America, Washington, USA}

\author{Nariaki V. Nitta}
\affiliation{Lockheed Martin Solar and Astrophysics Laboratory, Palo Alto, CA 94304 USA}

%%%%%%%%%%%%%%%%%%%%%%%%%%%%%%%%%%%%%%%%%%%%%%%%%%%%%%%%%%%%%%%%%%%%%%%%%%%%%%%%

\begin{abstract}

Gradual solar energetic particle (SEP) events, usually attributed to shock waves driven by coronal mass ejections (CMEs), show a wide variety of temporal behaviors. For example, TO, the $>$10~MeV proton onset time with respect to the launch of the CME, has a distribution of at least an order of magnitude, even when the source region is not far from the so-called well-connected longitudes.  It is important to understand what controls TO, especially in the context of space weather prediction. Here we study two SEP events from the western hemisphere that are different in TO on the basis of  $>$10~MeV proton data from the Geostationary Operations Environmental Satellite, despite %but 
similar in the CME speed and longitude of the source regions. We try to find the reasons for different TO, or proton release times, in how the CME-driven shock develops and the Alfv\'{e}n Mach number of the shock wave reaches some threshold, by combining the CME height-time profiles with radio dynamic spectra.  We also discuss how CME-CME interactions and active region properties may affect proton release times.

\end{abstract}

%%%%%%%%%%%%%%%%%%%%%%%%%%%%%%%%%%%%%%%%%%%%%%%%%%%%%%%%%%%%%%%%%%%%%%%%%%%%%%%%


%%%%%%%%%%%%%%%%%%%%%%%%%%%%%%%%%%%%%%%%%%%%%%%%%%%%%%%%%%%%%%%%%%%%%%%%%%%%%%%%

%%%%%%%%%%%%%%%%%%%%%%%%%%%%%%%%%%%%%%%%%%%%%%%%%%%%%%%%%%%%%%%%%%%%%%%%%%%%%%%%
% Introduction
%%%%%%%%%%%%%%%%%%%%%%%%%%%%%%%%%%%%%%%%%%%%%%%%%%%%%%%%%%%%%%%%%%%%%%%%%%%%%%%%
\section{Introduction} \label{sec:intro}

Gradual solar energetic particle (SEP) events are almost always accompanied by fast and extended coronal mass ejections (CMEs) that drive shock waves. These SEP events can be extremely intense, posing various space weather impacts, for example, on human bodies, satellite operations, high-frequency communications, etc. Their temporal variations as well as magnitudes are among the most important items of space weather prediction. There is a general trend of timescales with respect to the locations of the associated flares such that SEP events originating in regions in the western hemisphere start earlier and reach the peak fluxes in shorter times than those occurring elsewhere \citep{cane_1988_jgr}.

\citet{kahler_2005_apj} and \citet{kahler_2013_apj} introduced three timescales as follows,
TO: the SEP onset time with respect to the CME launch, TR: the rise time from the SEP onset time to the SEP half-peak during the rising phase, TD: the duration between SEP half-peak during the rising and declining phases.
These papers revealed that TR and TD were positively correlated with CME speed, and interpreted that fast CME continued to drive the shock wave and injected SEP for a long time. They also revealed that TO was related to CME speed and peak proton flux, but no correlation with the acceleration of CME was found.
TO is particularly challenging to understand, as we know of some events with short TO from likely far side regions that are almost certainly ill-connected \citep[e.g.,][]{cliver_2005_icrc, gomez_herrero_2015_apj, kahler_2016_apj}.  TO, as determined by first-arriving particles, may contain more information on acceleration processes close to the Sun than TR and TD, which may be more susceptible to transport processes.  

In a recent statistical study of the association of fast CMEs with SEP events mostly during solar cycle 24 \citep{kihara_2020_apj}, the three timescales were measured and compared with the source locations and CME speeds. In particular, TO was found to be short 
if the source region was within 60$\arcdeg$ in longitude from the footpoint of the Parker spiral (median: 86 minutes but 308 minutes in other longitudinal ranges), and negatively correlated with the CME speed for better connected events. But the scatter was quite large even for events with small longitudinal separations from the footpoint of the Parker spiral. 

In this paper, we further investigate two events from \cite{kihara_2020_apj} that apparently had different TO, despite their similar source locations in the western hemisphere and similar CME speeds of $\sim$1200~km~s$^{-1}$. We explore the possibility that
the event with longer TO may reflect a slow growth of the CME-driven shock wave that becomes strong enough for particle acceleration only at later times. Combining CME height-time profiles with radio dynamic spectra that contain type II radio bursts, we follow the temporal evolution of the Alfv\'{e}n Mach number of the shock wave
with time above the two active regions without conducting advanced modeling.  In Section~\ref{sec:overview}, we describe the event selection and give an overview of the two events. We revisit in Section~\ref{sec:more_sep} the SEP timescales that are used for the subsequent analysis. In Section~\ref{sec:analysis}, we study how the shock waves develop in the two events in relation to TO or the SEP release times.  In addition we study other factors that may affect these times.  We summarize our findings in Section~\ref{sec:summary}.

\section{Observations}
\label{sec:overview}

\subsection{Event Selection} \label{sec:selection}
\citet{kihara_2020_apj} conducted a statistical study of energetic CMEs that occurred between December 2006 and October 2017 in terms of their associations with SEP events.
They also studied the timescales of the associated SEP events with respect to the speeds and source locations of the CMEs as shown in the Table 2 of \citet{kihara_2020_apj}. They based the SEP analysis on data from the Energetic Particle Sensor \citep{onsager_1996_spie} on the Geostationary Operations Environmental Satellite (GOES), and the High-Energy Telescope \citep[HET;][]{von_rosenvinge_2008_ssr} and the Low-Energy Telescope \citep[LET;][]{mewaldt_2008_ssr}, which belong to the suite of instruments for the In Situ Measurements of Particles and CME Transients \citep[IMPACT;][]{luhmann_2008_ssr} on the Solar-Terrestrial Relations Observatory \citep[STEREO;][]{kaiser_2008_ssrv}. The SEP events were identified when the $>$10~MeV proton flux exceeded 1 particle flux unit (pfu; defined as particles s$^{-1}$ sr$^{-1}$ cm$^{-2}$). The CMEs responsible for the SEP events and the associated flares were found in white-light coronagraph and EUV low-coronal images produced by the instruments on the Solar and Heliospheric Observatory \citep[SOHO;][]{domingo_1995_sp}, Solar Dynamics Observatory \citep[SDO;][]{pesnell_2012_sp}, and STEREO. 
As expected, \cite{kihara_2020_apj} found that SEP events that occurred in regions not far from the magnetic footpoints of the observer tend to have shorter timescales (in both TO and TR, see Figure~6 of the paper). 
However, TO mostly (77/82) ranges from 0.5 to 4 hours even when the longitudinal separation of the region from the Parker spiral footpoint is less than 60$\arcdeg$.  TO also appears to depend on the speed of the associated CME. 

%%%%%%%%%%%%%%%%%%%%%%%%%%%%%%%%%%%%%%%%
%%%%%%%%%%%%%%%%%%%%%%%%%%%%%%%%%%%%%%%%
\begin{deluxetable*}{lllrrrlrrlrrl}[h]
\label{table}
\tablenum{1}
\tabletypesize{\scriptsize}
\tablecaption{Basic Parameters of the Two SEP Events}
\tablehead{\colhead{} & \colhead{} & \multicolumn{4}{c}{CME} & \colhead{} & \multicolumn{2}{c}{type II radio burst\tablenotemark{e}} & \colhead{} & \multicolumn{2}{c}{SEP event} \\
\cline{3-6} \cline{8-9} \cline{11-12}
\colhead{ID} & \colhead{} & \colhead{launch\tablenotemark{a}} & 
\colhead{speed\tablenotemark{b}} & 
\colhead{width\tablenotemark{c}} & 
\colhead{source\tablenotemark{d}} & 
\colhead{} & \colhead{frequency} & \colhead{time} & \colhead{} &
\colhead{$I_{p}$\tablenotemark{f}} & \colhead{TO\tablenotemark{g}}\\
\colhead{} & \colhead{} & \colhead{date and time} & \colhead{(km s$^{-1}$)}  & \colhead{(deg)} & \colhead{location} & \colhead{} & \colhead{(MHz)} & \colhead{} & \colhead{} & \colhead{(pfu)} & \colhead{(min)}\\
}
\startdata
Event 1 & & 2014-04-18 12:43 & 1203 & 360 & S20W34 & & 60 & 12:55 & & 58.5 & 62\\
Event 2 & & 2017-07-14 01:12 & 1200 & 360 & S06W29 & & 14 & 01:20 & & 13.6 & 158\\
\enddata
\tablenotetext{\tiny a}{The launch time of CME calculated by extrapolating the height-time relations from the LASCO C2 and C3 data to the solar surface. Cited from the LASCO CME catalog.}
\tablenotetext{\tiny b}{The linear speed obtained by fitting %a straight line with 
whole data points in LASCO C2 and C3. Cited from the LASCO CME catalog.}
\tablenotetext{\tiny c}{The width in the plane of the sky of CME measured in LASCO C2 FOV. Cited from the LASCO CME catalog.}
\tablenotetext{\tiny d}{The location of %source active region of 
associated flare analyzed in \cite{kihara_2020_apj}.}
\tablenotetext{\tiny e}{Frequency and time at the onset of the associated type II radio burst in each event.}
\tablenotetext{\tiny f}{%Peak $>$10~MeV proton flux 
The peak proton flux with energies above 10 MeV observed by GOES satellite. Defined in \cite{kihara_2020_apj}.}
\tablenotetext{\tiny g}{The $>$10~MeV proton onset time with respect to the launch of the CME. Defined in \cite{kihara_2020_apj}.}

\end{deluxetable*}

%%%%%%%%%%%%%%%%%%%%%%%%%%%%%%%%%%%%%%%%
%%%%%%%%%%%%%%%%%%%%%%%%%%%%%%%%%%%%%%%%

In this paper, we selected two events that have widely different TO (i.e., 62 and 158 minutes) even though they came from regions in similar longitudes and were associated with halo CMEs with similar speeds. They occurred on 2014 April 18 and 2017 July 14. We hereafter refer to these SEP events as Event~1 and Event~2, respectively. Their basic parameters are shown in Table~\ref{table}. The primary purpose of this work is to explain this wide difference in TO.
We also revise the SEP onset times in Section~\ref{sec:more_sep}, which we will use in the subsequent analyses.

\subsection{Overview of the Events} \label{overview}

%%%%%%%%%%%%%%%%%%%%%%%%%%%%%%%%%%%%%%%%
%%%%%%%%%%%%%%%%%%%%%%%%%%%%%%%%%%%%%%%%
\begin{figure*}[h]
\gridline{\fig{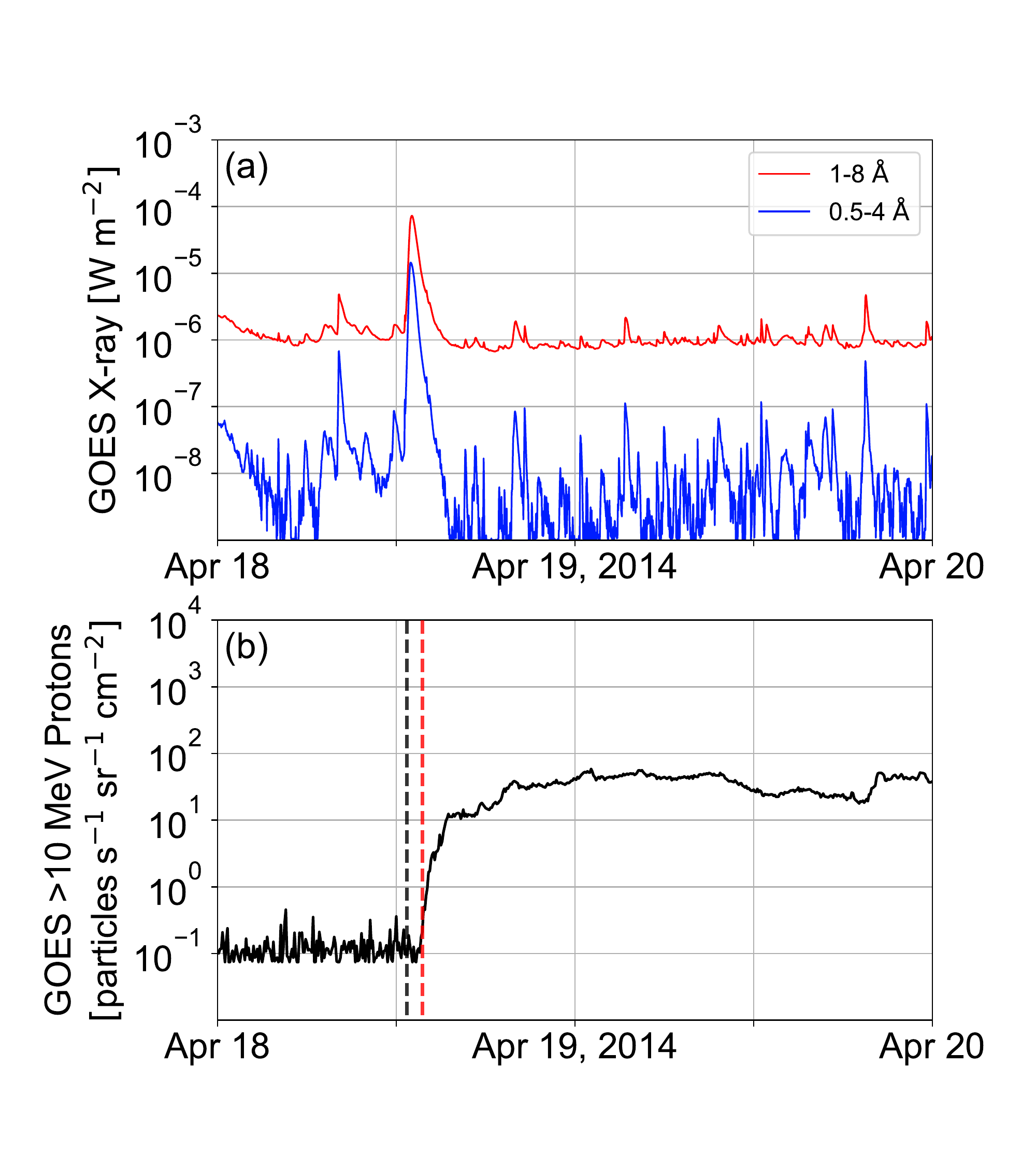}{0.49\textwidth}{}
          \fig{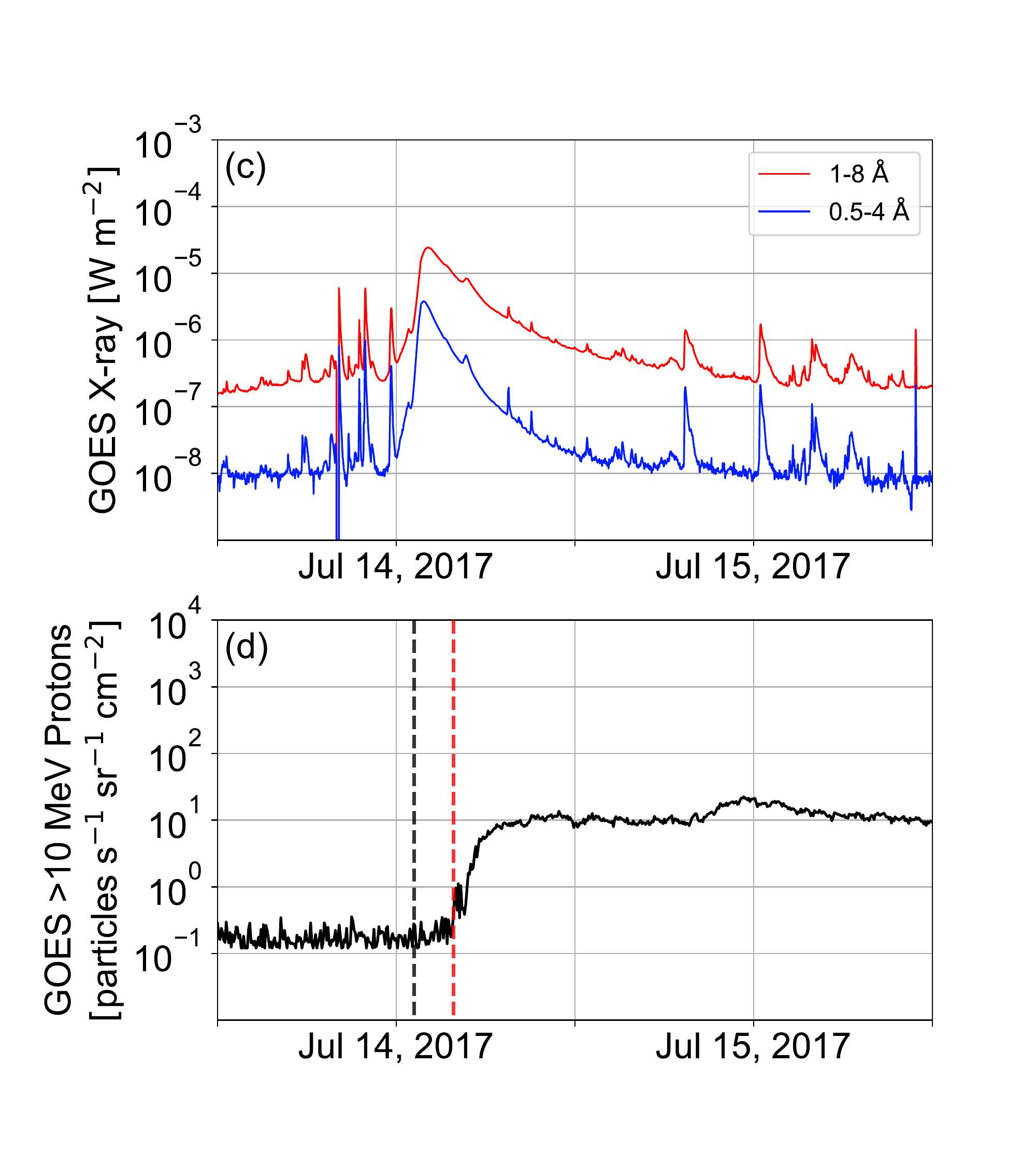}{0.49\textwidth}{}}
\caption{The soft X-ray (SXR) and the integrated flux of $>$10~MeV protons observed by GOES satellite for Event~1 ((a) and (b)) and for Event~2 ((c) and (d)). In panels (a) and (c), red and blue lines correspond to 1-8 $\mathrm{\mathring{A}}$ and 0.5-4 $\mathrm{\mathring{A}}$. In panels (b) and (d), the launch time of CME and the time of proton onset are indicated by the black and red dashed lines, respectively.
}
\label{fig:goes_xp}
\end{figure*}
%%%%%%%%%%%%%%%%%%%%%%%%%%%%%%%%%%%%%%%%
%%%%%%%%%%%%%%%%%%%%%%%%%%%%%%%%%%%%%%%%

In Figure~\ref{fig:goes_xp} we plot the soft X-ray (SXR) and SEP (proton) time profiles of the two events over two-day intervals. The flare associated with Event~1 (in panel (a)) is M7.3 in the GOES classification (the peak 1\,--\,8~\AA\ flux of 7.3$\times$10$^{-5}$~W m$^{-2}$), whereas the one associated with Event~2 (in panel (c)) is M2.4. The latter flare is of much longer duration, staying above the pre-event level in the GOES 1\,--\,8~\AA\ channel for more than two days. Both flares are associated with halo CMEs, whose mean linear speed is $\sim$1200~km~s$^{-1}$ across the combined field of view (FOV) of the C2 and C3 telescopes of the Large Angle Spectrometric Coronagraph \citep[LASCO;][]{brueckner_1995_sp} on board SOHO. The CME launch times in black dashed lines are calculated by extrapolating the height-time relations from the LASCO C2 and C3 data to the unit height (1 solar radius~R$_{\sun}$), i.e. the solar surface. 

%%%%%%%%%%%%%%%%%%%%%%%%%%%%%%%%%%%%%%%%
%%%%%%%%%%%%%%%%%%%%%%%%%%%%%%%%%%%%%%%%
\begin{figure*}[h]
\gridline{\fig{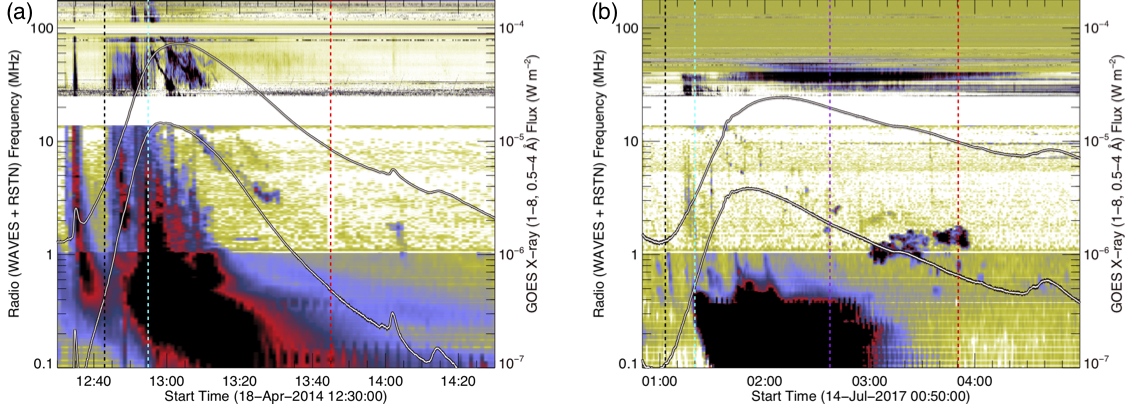}{0.99\textwidth}{}}
\caption{Radio dynamic spectra of the two events in the combined metric and DH ranges.  The latter data are obtained with the Wind/WAVES instrument, and the metric data obtained at (a) RSTN/San Vito for Event~1 and (b) Culgoora Observatories for Event~2, respectively. The black (red) dashed lines indicate the CME launch times (the onset times of $>$10~MeV protons as observed by GOES), which replicate those in Figure~\ref{fig:goes_xp}.  The cyan dashed lines indicate the start times of the type II radio bursts.  The purple line in (b) indicates a proton onset time from SOHO/ERNE (see Section~\ref{sec:more_sep}).
}
\label{fig:radio}
\end{figure*}
%%%%%%%%%%%%%%%%%%%%%%%%%%%%%%%%%%%%%%%%
%%%%%%%%%%%%%%%%%%%%%%%%%%%%%%%%%%%%%%%%

Both Event~1 and Event~2 are accompanied by type II radio bursts, while
their appearances are quite different as found in Figure~\ref{fig:radio}, where we show radio dynamic spectra between 180~MHz and 0.1~MHz that consist of data from ground-based observatories and the Radio and Plasma Wave Experiment \citep[WAVES;][]{bougeret_1995_ssr} on the Wind spacecraft. In Event~1 (Figure~\ref{fig:radio}(a)), the type II radio burst started at 12:55~UT from about 60~MHz (fundamental), which is 12~minutes after the CME launch (12:43~UT) and 8~minutes before the SXR peak (13:03~UT). It is preceded by strong type III radio bursts during the flare impulsive phase. 

In Event~2 (Figure~\ref{fig:radio}(b)), the type II radio burst is weak and intermittent
and seen only in Wind/WAVES data below 14~MHz. It started at 01:20~UT, which is 12~minutes after the CME launch (01:12~UT) and 49~minutes before the SXR peak (02:09~UT). 
Type III radio bursts are also weak in Event~2,
mostly at frequencies below the type II radio burst, sometimes categorized as shock-accelerated 
events \citep{cane_1981_grl}.  Although type II radio bursts are widely considered to signify shock waves, accelerating $\lesssim$10~keV electrons, the proton onset is delayed in Event~2 much more than expected of $\sim$10~MeV protons, as reflected in larger TO.  Lastly, note strong emissions starting around 03:00~UT in Figure~\ref{fig:radio}(b). They do not follow the frequency drift of the type II radio burst.  These features may indicate an interaction of the CME in Event~2 with a previously-launched CME \citep{gopalswamy_2002_apj}. In Section~\ref{sec:interaction} we will briefly discuss the possible effect of this CME-CME interaction on the observed SEPs in Event~2. 

%%%%%%%%%%%%%%%%%%%%%%%%%%%%%%%%%%%%%%%%
%%%%%%%%%%%%%%%%%%%%%%%%%%%%%%%%%%%%%%%%
\begin{figure*}[h]
\gridline{\fig{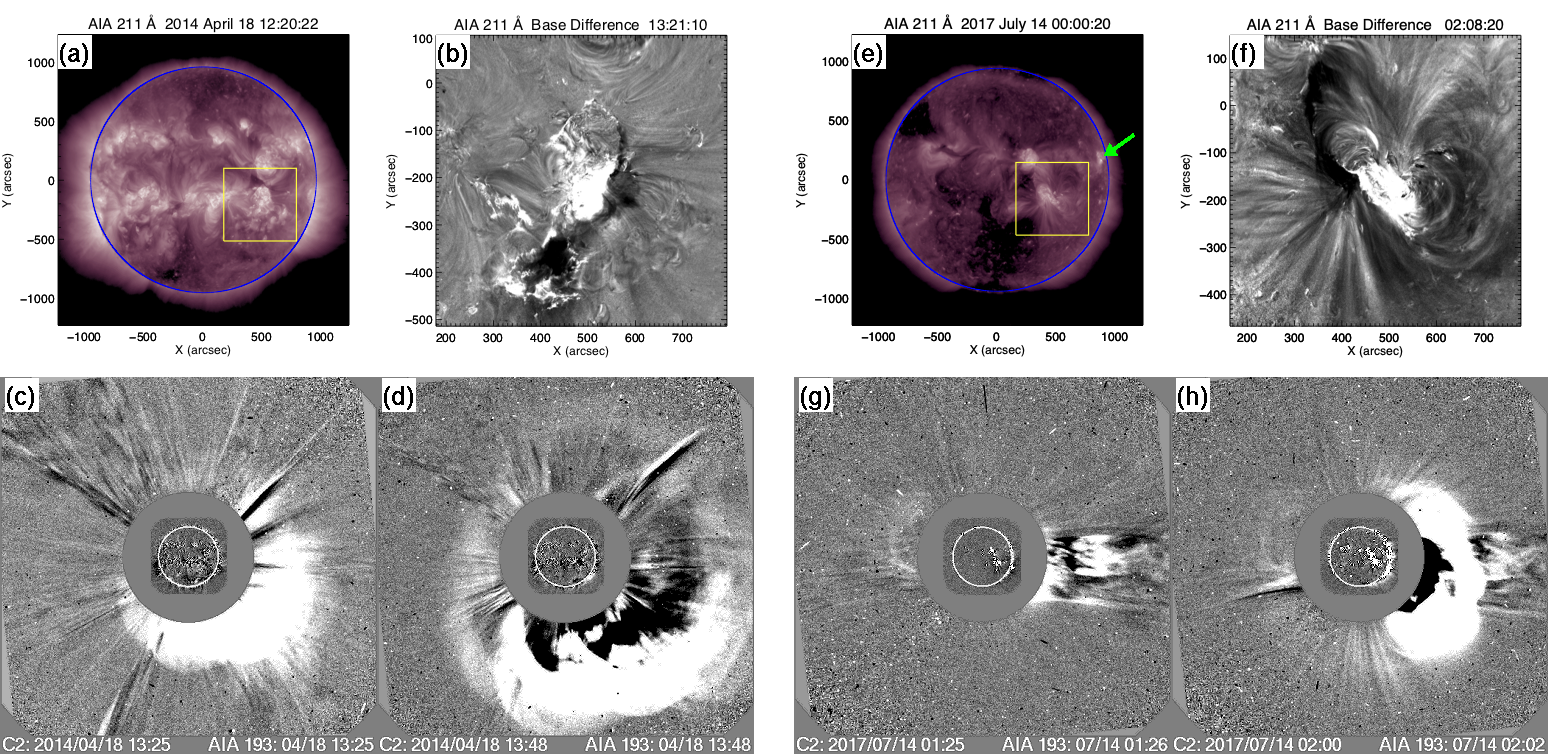}{0.99\textwidth}{}}
\caption{Low coronal and coronagraph images for Event~1 ((a)\,--\,(d)) and Event~2 ((e)\,--\,(h)). (a) and (e): AIA 211~\AA\ images prior to the eruptions that led to the CMEs. The active regions that hosted the eruptions are included in the yellow boxes, in which coronal dimmings are noted in difference images with the pre-eruption image subtracted ((b) and (f)).  (c) and (g): first available LASCO images of the CMEs in Event~1 and Event~2. (d) and (h): later LASCO images. The green arrow in (e) points to AR~12667, which produced narrower and slower CMEs than that in Event~2.}
\label{fig:aia_lasco}
\end{figure*}
%%%%%%%%%%%%%%%%%%%%%%%%%%%%%%%%%%%%%%%%
%%%%%%%%%%%%%%%%%%%%%%%%%%%%%%%%%%%%%%%%

Spatially-resolved coronal observations of the two events are given in Figure~\ref{fig:aia_lasco}, where panels (a)\,--\,(d) and (e)\,--\,(h) cover Event~1 and Event~2, respectively. Low coronal images (panels (a), (b), (e), and (f)) come from 211~\AA\ channel of the Atmospheric Imaging Assembly \citep[AIA;][]{lemen_2012_sp} on board SDO. The remaining panels consist of coronagraph images that come from LASCO. The origins of the eruptions\,--\,NOAA AR~12036 at S17W35 for Event~1 and NOAA AR~12665 at S06W29 for Event~2\,--\,are contained in the yellow boxes in Figures~\ref{fig:aia_lasco}(a) and \ref{fig:aia_lasco}(d), in which we note coronal dimmings in pre-event subtracted images (Figures~\ref{fig:aia_lasco}(b) and \ref{fig:aia_lasco}(f)).  Both events are associated with a halo CME, although asymmetric, as seen in Figures~\ref{fig:aia_lasco}(d) and \ref{fig:aia_lasco}(h). 

Figures~\ref{fig:aia_lasco}(c) and \ref{fig:aia_lasco}(g) show the first available LASCO C2 images of the CMEs in the two events.  It appears that we miss an early development of the CME in Event~1 due to the data gap of $\sim$40~minute preceding the image in Figure~\ref{fig:aia_lasco}(c). The CME in Event~2 was preceded by a narrower CME, which was associated with a C3.0 flare from AR~12667 around N12W71. This region, indicated by a green arrow in Figure~\ref{fig:aia_lasco}(e), produced C2.0, C5.9, and C3.0 flares starting, respectively, at 21:27, 21:46, and 23:30~UT on July 13. All of them produced a slow and narrow CME, and an electron event across the 10~keV\,--\,2~MeV range but not a proton event. When protons increased in Event~2, the electron background was still elevated due to the electron event associated with the C3.0 flare, so it is not clear whether Event~2 produced an electron event. In contrast, Event~1 was accompanied by a strong electron event well above the elevated background in Event~2. The CME in Event~2 apparently caught up with the narrow CME and possibly resulted in a CME-CME interaction suggested in Figure~\ref{fig:aia_lasco}(h). However, this is an hour earlier than the CME-CME interaction indicated in radio data (Figure~\ref{fig:radio}(b)).

\section{Further Analysis of SEP Events}
\label{sec:more_sep}

%%%%%%%%%%%%%%%%%%%%%%%%%%%%%%%%%%%%%%%%
%%%%%%%%%%%%%%%%%%%%%%%%%%%%%%%%%%%%%%%%
\begin{figure*}
\gridline{\fig{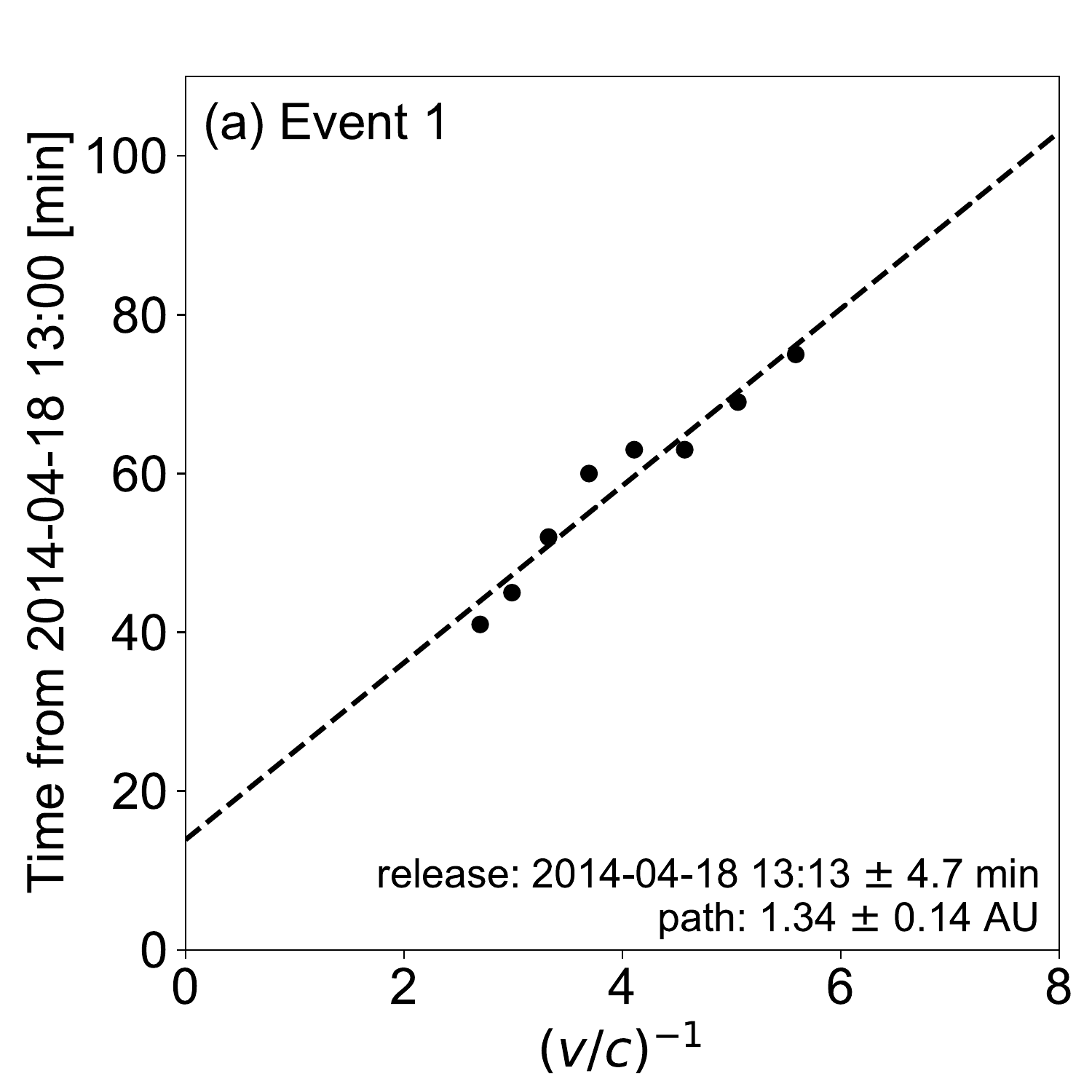}{0.49\textwidth}{}
          \fig{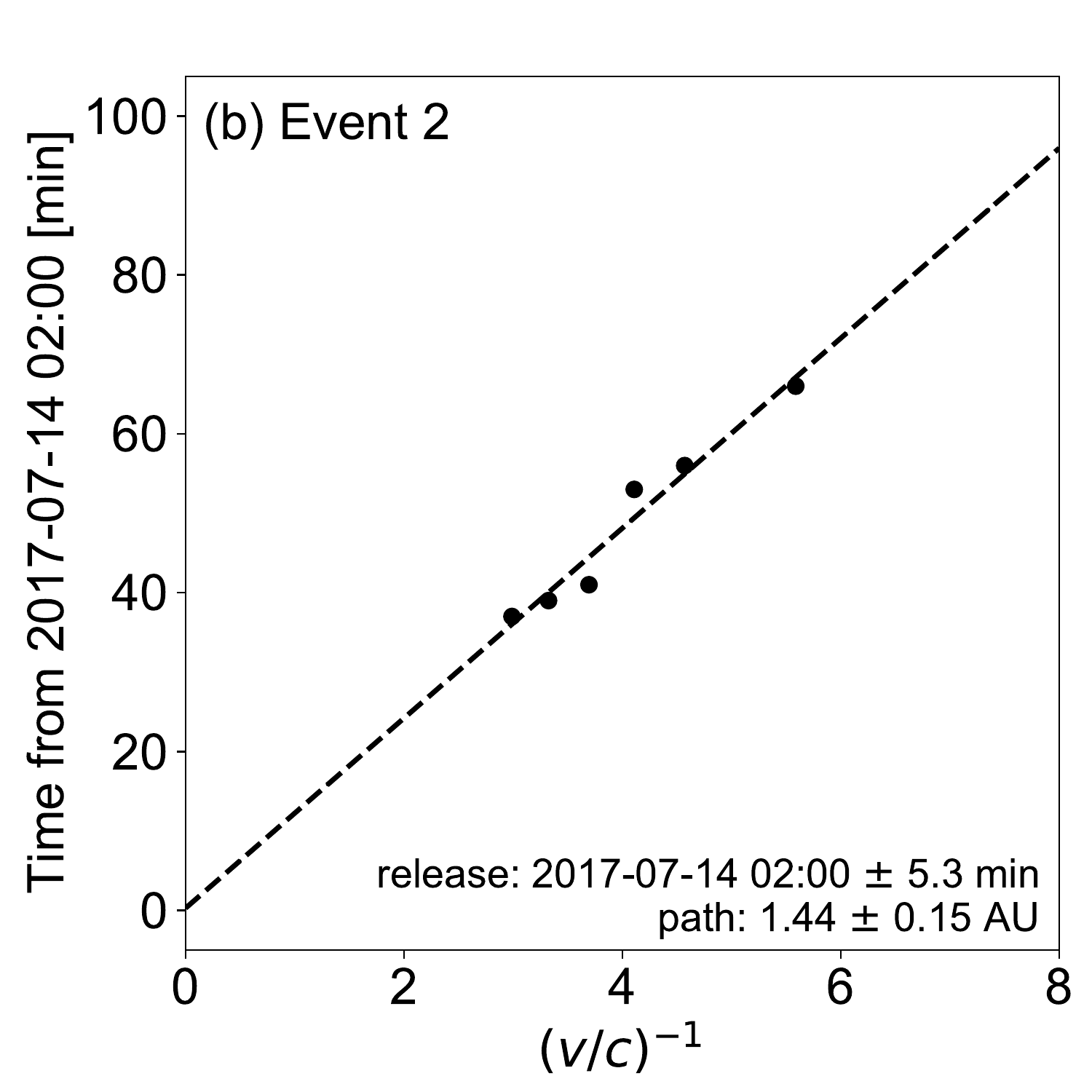}{0.49\textwidth}{}}
\caption{VDA analysis based on ERNE/HED data.  The observed onset times are plotted against the inverse velocities $(v/c)^{-1}$ calculated from the effective energies of the individual channels. The vertical axis is the elapsed time since (a) 2014 Apr 18 13:00 and (b) 2017 July 14 2:00, respectively.
}
\label{fig:vda}
\end{figure*}
%%%%%%%%%%%%%%%%%%%%%%%%%%%%%%%%%%%%%%%%
%%%%%%%%%%%%%%%%%%%%%%%%%%%%%%%%%%%%%%%%

Here, we re-evaluate TO of the two events.
The GOES energetic particle data suffer from high background, which may prevent the SEP onset from being properly captured if the particle flux rises slowly from a low level.  Another problem may be a possibly inadequate energy discrimination because the detector is only passively shielded \citep{posner_2007_sw, kuhl_2019_sw}.  These issues drive us to study similar data from other instruments.  Here we analyze data from the High Energy Detector (HED) of the Energetic and Relativistic Nuclei and Electron \citep[ERNE;][]{torsti_1995_sp} on board SOHO, which measures protons in the energy range of 13\,--\,130~MeV divided into 10 channels and has much lower background.  In Event~1, protons were detected above the background up to the 64\,--\,80~MeV channel, where the onset time is found to be the same ($\pm$5 ~minutes) as that of the GOES $>$10~MeV integral channel; the onset time of the 13\,--\,16~MeV channel comes $\sim$30 minutes later.  We keep the same TO that was calculated by \cite{kihara_2020_apj} for Event~1, noting that it refers to the ERNE 64\,--\,80~MeV channel.  
In Event~2, on the other hand, the onset times of all the ERNE/HED channels that detected protons above the background (up to the 50\,--\,64~MeV channel) are much earlier than that of the GOES $>$10~MeV integral channel, suggestive of an effect of the high background of the latter.  The ``revised'' onset time, coming from the 50\,--\,64~MeV channel of ERNE/HED, is used to redefine TO as indicated by the purple line in Figure~\ref{fig:radio}(b).  This gives TO$ =$85  minutes (down from 158 minutes). 
The updated TO for Event~2 is still $\sim$25~minutes longer than TO for Event~1.

%%%%%%%%%%%%%%%%%%%%%%%%%%%%%%%%%%%%%%%%
%%%%%%%%%%%%%%%%%%%%%%%%%%%%%%%%%%%%%%%%
\begin{figure*}
\gridline{\fig{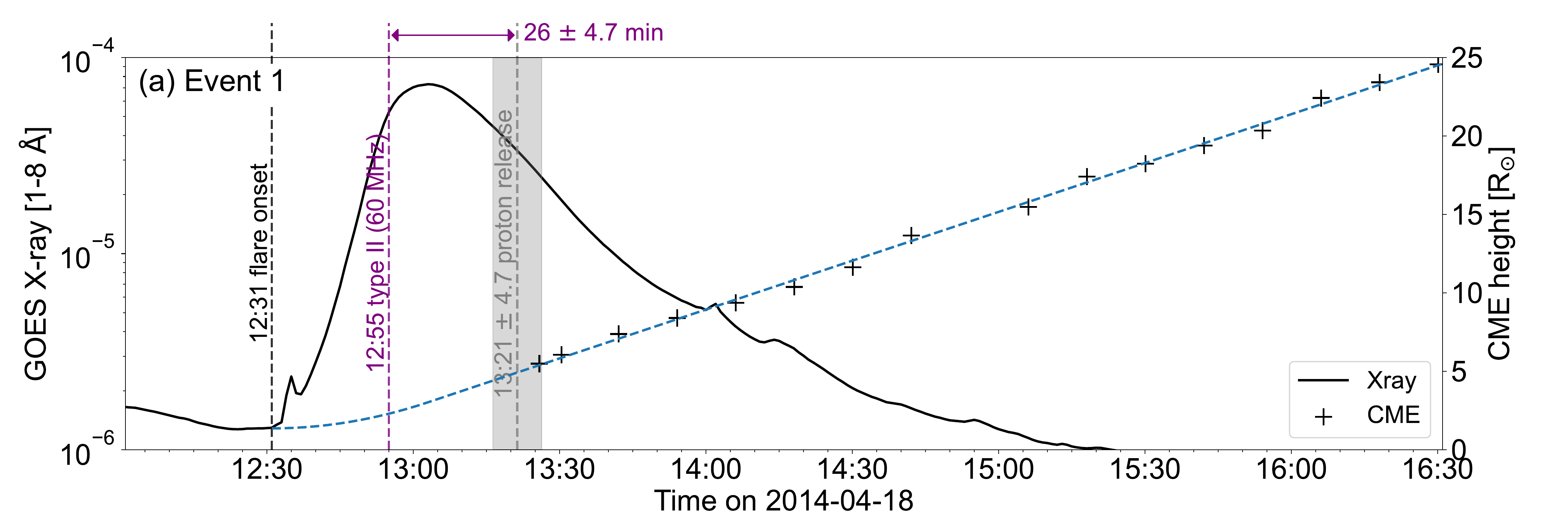}{0.98\textwidth}{}}
\gridline{\fig{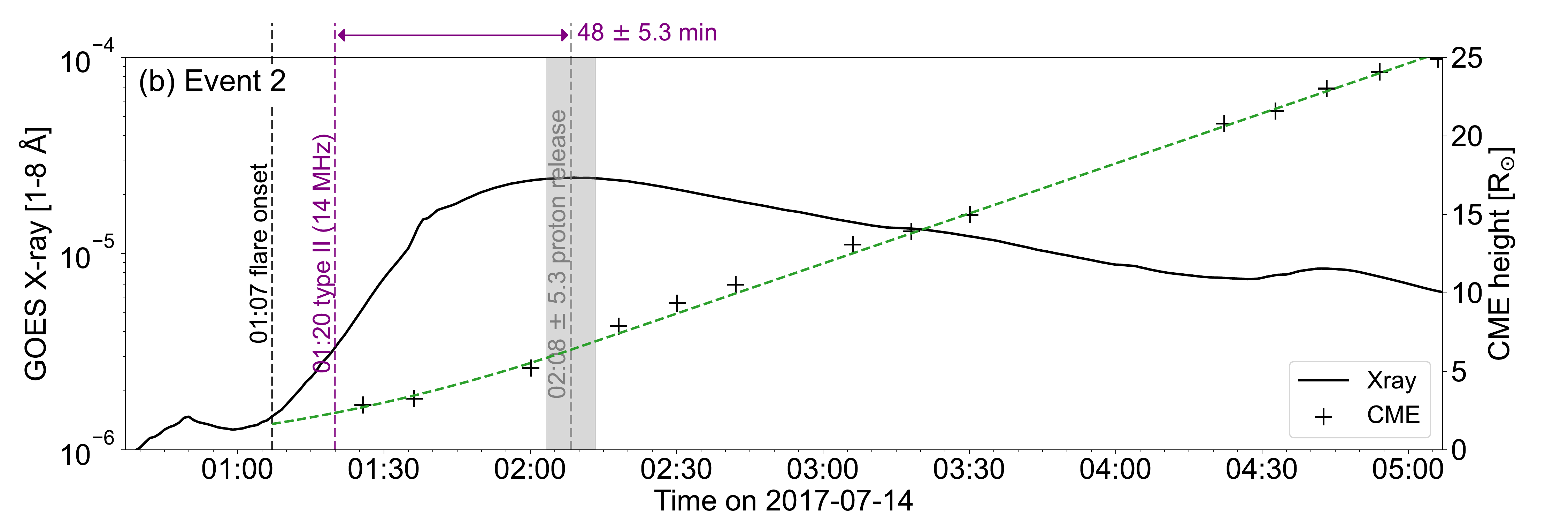}{0.98\textwidth}{}}
\caption{Summary of the timeline for (a) Event~1 and (b) Event~2. SXR (1\,--\,8$\mathrm{\mathring{A}}$) flux observed by the GOES satellite is shown as solid curves (black). The height of the leading edge of the CME is shown as crosses (measurements) and dashed curves (models). The vertical dashed lines in black and purple indicate the onset times of the flare and type II radio burst, respectively. The shaded areas in red indicates the proton release times with uncertainties (described in Section \ref{sec:more_sep}). The purple arrow indicates the interval between the type II onset and the estimated proton release from the VDA analysis. 
}
\label{fig:timeline}
\end{figure*}
%%%%%%%%%%%%%%%%%%%%%%%%%%%%%%%%%%%%%%%%
%%%%%%%%%%%%%%%%%%%%%%%%%%%%%%%%%%%%%%%%

We also conduct the Velocity Dispersion Analysis \citep[VDA; see, e.g.,][]{vainio_2013_swsc}, using all the ERNE/HED channels in which protons were detected above the background (up to the 64\,--\,80~MeV channel for Event~1 and the 50\,--\,64~MeV channel for Event~2).  We plot the onset times (per visual inspection) against the inverse of speed $(v/c)^{-1}$, which corresponds to the effective energy of the channel (Figure~\ref{fig:vda}). A least-square fit yields the proton release time of 2014 April 18 13:13~UT$\pm$4.7 minutes and 2017 July 14 02:00~UT$\pm$5.3 minutes for Event~1 and Event~2, respectively. The associated path lengths come out as 1.32$\pm$0.14~AU and 1.44$\pm$0.15~AU, which are somewhat longer than the lengths of the nominal Parker Spiral for the observed solar wind speeds but within a range that suggests no major effect of scattering in the interplanetary space.
Figure~\ref{fig:timeline} shows the summary of the timeline of each event. We indicate the proton release times with 8.3~minutes added to account for the 1~AU travel time of light so that we can compare them with other electromagnetic-wave-based observations.
From now on, instead of TO, we shall investigate the proton release time corrected for the 1~AU travel time of light, even though we originally aimed at explaining different TO. Moreover, we concern the proton release time with respect to the start time of the type II radio burst rather than the CME launch time. The proton release is delayed by 26$\pm$4.7 minutes for Event~1 and 48$\pm$5.3 minutes for Event~2. So the difference still exists between Event~1 and Event~2, although not as large as in the original TO.  

Lastly, the multi-channel data from ERNE let us obtain fluence energy spectra of the two events. We integrate the background-subtracted proton flux in each channel while it is above the background. The spectral fitting gives the power-law index of 3.65 for Event~1 and 4.18 for Event~2.
These are close to the value of 3.83, that is the averaged indices of fluence spectra of well-connected SEP events reported by \cite{gopalswamy_2016_apj}. The slightly softer index in Event~2 may be an indication of a weaker shock, possibly related to a longer delay of the SEP onset time, but the difference may not be large enough to be conclusive.

\section{Factors that may control the particle release time}
\label{sec:analysis}

As shown in Section \ref{sec:more_sep}, protons are not released immediately after the formation of the shock wave as manifested in type II radio bursts. However, the time difference is longer for Event~2.  What is the reason for varying proton release times?  In the following, we consider the evolution of the CME-driven shock wave, CME-CME interaction and properties of the active region that produces the CME.

\subsection{Evolution of Shock Waves with Height}
\label{sec:shock}

In this section, we investigate the possibility that particles (protons) are accelerated and released only when the shock wave becomes strong enough. Specifically, we study how the Alfv\'{e}n Mach number ($M_A$) of the shock wave changes with time. The Alfv\'{e}n Mach number is  expressed as $M_A = (v_s - v_{sw})/v_A$, where $v_s$ is the shock speed, $v_{sw}$ is the solar wind speed, and $v_A$ is the Alfv\'{e}n speed.

\begin{figure*}
\gridline{\fig{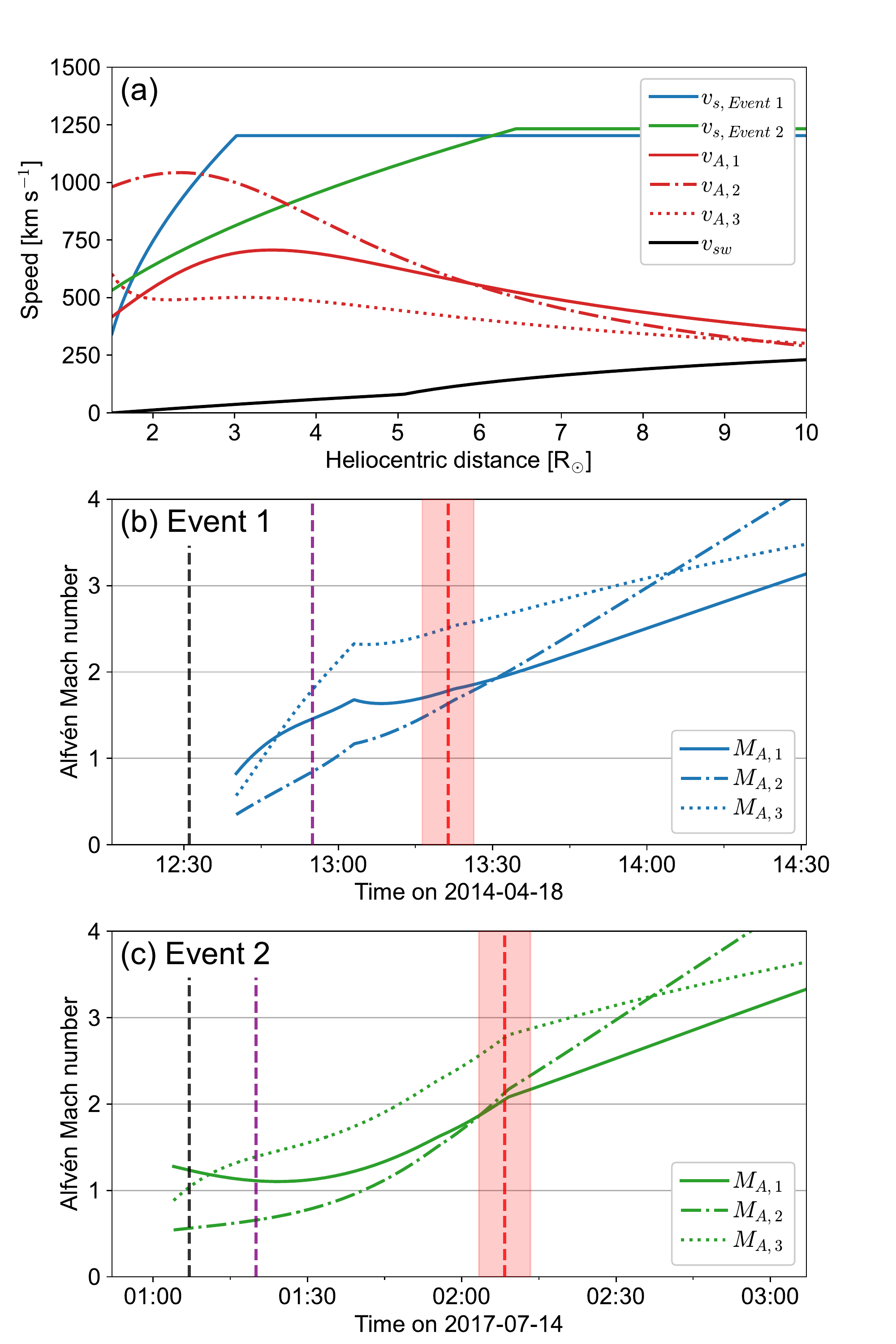}{0.5\textwidth}{}}
\caption{(a): Shock speed for Event~1 (blue, solid) and Event~2 (green, solid) calculated from the modeled height-time profiles. The red lines represent the Alfv\'{e}n speeds based on the three models of magnetic field. $v_{sw}$ is shown as the black solid line. (b) and (c): $M_{A}$ for Event~1 and Event~2, respectively, calculated with all the information presented in (a). As in Figure~\ref{fig:timeline}, the dashed lines in black and purple indicate the onset times of the flare and type II radio burst, respectively, and the shaded areas in red the proton release times with uncertainties.
}
\label{fig:mach}
\end{figure*}

For the shock speed, we could simply use the linear or quadratic fits to the height of the leading edge of the CME, as published in the CDAW LASCO CME catalog\footnote{\url{https://cdaw.gsfc.nasa.gov/CME\_list/}} \citep{yashiro_2004_jgr}. However, these fits are made on the height measurements in the whole (C2 and C3) FOV and may be too coarse to discuss the CME kinematics in the height range 
that likely corresponds to the SEP onset (e.g., below 10~R$_{\odot}$). Here we instead model the height-time profiles of CMEs such that they undergo constant acceleration from the onset to the peak of the SXR flux.  This may be justified by the general tendency of CMEs to accelerate in the flare impulsive phase \citep[e.g.,][]{zhang_apj_2004, temmer_2010_apj}. We further assume that CMEs move with a constant speed in the LASCO FOV after the SXR peak. 
This modeled CME height-time profile is meant to better reproduce the behavior of the shock speed near the proton release time, and does not necessarily match the information from the CME catalog, including the estimated time of CME launch.
The dashed curves in Figures~\ref{fig:timeline}(a) and \ref{fig:timeline}(b) show the modeled CME height-time profiles of Event~1 (blue) and Event~2 (green), respectively. For each event the shock speed $v_s$ is calculated using the modeled CME height-time profile (Figure~\ref{fig:mach}(a)). Even though the average speeds in the LASCO FOV are similar in both CMEs, their height-speed profiles are very different. The Event~1 CME (blue) accelerates quickly with a large acceleration of 627~m~s$^{-2}$ and reaches $\sim$1200~km~s$^{-1}$ before 3~$R_{\odot}$, while the Event~2 CME (green) accelerates slowly (188~m~s$^{-2}$) and reaches $\sim$1300~km~s$^{-1}$ at 6.6~$R_{\odot}$. 

The Alfv\'{e}n speed, $v_A$, depends on the density and magnetic field, neither of which is directly observed, so we must rely on models. In order to address the inherently model-dependent nature of our attempt to calculate the Alfv\'{e}n Mach number of the shock waves, we use the frequency drift of the type II radio burst to constrain the density profile with height. We choose the density model that places the shock wave of the type II radio burst at heights closest to the modeled CME heights at overlapping times. Consider the frequency (fundamental) of the type II radio burst to be the local plasma frequency, and we can get the density. For both events, the 3-fold (multiplied by 3) Saito model \citep{Saito_1977_sp} yields the heights of the shock wave of the type II burst that best match the modeled CME heights. For the magnetic field, we consider the following three models:
\begin{eqnarray}
  B_1(r) &=& 2.2r^{-2} \\
  B_2(r) &=& 6r^{-3}+1.18r^{-2} \\
  B_3(r) &=& 0.5(r-1)^{-1.5}
\end{eqnarray}
These are (1) the model assuming magnetic flux conservation\citep{mann_1999_proc}, (2) the model based on measurement of Faraday rotation \citep{patzold_1987_sp}, and (3) the empirical model by \citet{dulk_mclean_1978_sp}.
Three Alfv\'{e}n speed profiles derived from these three magnetic field models ($B_1$, $B_2$, and $B_3$) are shown in the red solid ($v_{A,1}$),  dash-dotted ($v_{A,2}$), and dotted lines ($v_{A,3}$) in Figure~\ref{fig:mach}(a), respectively.

For the solar wind speed, $v_{sw}$, the model by \cite{sheeley_1997_apj} has been widely used, but it starts only at 4.5~R$_{\odot}$. Recently, $v_{sw}$ closer to the Sun (down to 1.53~R$_{\odot}$) has been obtained by \citet{bemporad_2021_aa}. We use the latter model up to the height of 5.07~R$_{\odot}$ (where $v_{sw}$ from the former model becomes larger), and the former model at greater heights. 
The solar wind speed profile is shown as the black line in Figure~\ref{fig:mach}(a).

We finally calculate three versions of $M_A$, based on the three magnetic field models $B_1$, $B_2$, and $B_3$ that were used to calculate $v_{A}$.  Figure~\ref{fig:mach}(b) shows the evolution of $M_{A}$ with time for Event~1. The different types of lines for $M_{A,1}$, $M_{A,2}$, and $M_{A,3}$, distinguish the corresponding magnetic field models, $B_1$, $B_2$, and $B_3$. The vertical lines and the shaded area are identical to those in Figure~\ref{fig:timeline}. Figure~\ref{fig:mach}(c) is the same as Figure~\ref{fig:mach}(b) but for Event~2. Note that the reliability of the magnetic field models may be somewhat compromised near the solar surface.  For example, the model of $B_2$ was originally calculated only in the range of 2\,--\,15~$R_{\odot}$. Accordingly, Figure~\ref{fig:mach} show the result only in $>1.5~R_{\odot}$. 

Despite an apparent dependence of $M_A$ on the assumed magnetic field models, we may understand the proton release times in relation to $M_A$. In both cases, all the three models of magnetic field yield $M_A$ that increase toward the proton release times. Including errors, $M_A$ reaches 1.6\,--\,2.6 for Event~1 and 2.0\,--\,3.0 for Event~2 during the estimated proton release time. Although it is beyond the scope of our work to discuss the critical Mach number \citep[e.g.,][]{bemporad_2011_apjl, rouillard_2016_apj}, $M_A$ in the above ranges may serve as thresholds, above which protons can be accelerated.
 
In Event~1, when the CME ceases to accelerate at $\sim$3~$R_{\odot}$ and $\sim$10 minutes after the onset of the type II radio burst, the shock speed already reaches $\sim$1200~km~s$^{-1}$. However, the Alfv\'{e}n Mach number remains low, because of the high Alfv\'{e}n speed due to strong magnetic field at a low altitude. $M_A$ reaches the critical value only $\sim$20~minutes after the CME stops accelerating as $v_A$ decreases. In Event~2, on the other hand, the CME continues to accelerate for $\sim$50~minutes after the onset of the type II radio burst until it travels to the height of $\sim$6.6~$R_{\odot}$. During the acceleration phase, $M_A$ only slowly increases, until it reaches the threshold as the CME attains the speed of $\sim $1200~km~s$^{-1}$.  This may explain why the proton release is delayed more in Event~2 than in Event~1.  It also aligns with a longer duration of the flare in Event~2. 

%%%%%%%%%%%%%%%%%%%%%%%%%%%%%%%%%%%%%%%%
%%%%%%%%%%%%%%%%%%%%%%%%%%%%%%%%%%%%%%%%

%%%%%%%%%%%%%% memo %%%%%%%%%%%%%%%%
%% Event 1:
%% CME launch: 12:43, type II: 12:55, flare peak: 13:03, proton release: 13:21 +/- 5, GOES X-ray flare onset 12:31
%% Event 2:
%% CME launch: 01:12, type II: 01:20, flare peak: 02:09, proton release: 02:08 +/- 5, GOES X-ray flare onset 01:07
%%%%%%%%%%%%%%%%%%%%%%%%%%%%%%%%%%%%

\subsection{CME-CME Interaction}
\label{sec:interaction}

As an alternative explanation for a later particle release in Event~2, let us assume that the shock is in fact too weak for particle acceleration, irrespective of the analysis given in Section~\ref{sec:shock}. Then what distinguishes Event~2 is the CME-CME interaction, which may compensate for the weak shock. It is proposed that when a fast CME catches up with a preceding CME, preconditioning by the preceding CME results in efficient particle acceleration \citep[e.g.,][]{gopalswamy_2002_apj, li_2005_icrc, li_2012_ssrv}.
The calculated proton release time is around 02:00~UT (Section \ref{sec:more_sep}), which is close to the time the CME in Event~2 caught up with the preceding narrow CME associated with a C3.0 flare in AR~12667 (Figure~\ref{fig:aia_lasco}(h)).  Note that the radio signatures that may indicate a CME-CME interaction starts only around 03:00~UT (Figure~\ref{fig:radio}(b)).  However, it is not clearly understood at which timing during a CME-CME interaction such signatures appear in radio spectra.

The possibility that Event~2 originally produced a weak shock wave with poor acceleration efficiency may be supported by the bandwidth of the type II radio burst. \citet{iwai_2020_apj} reported a positive correlation between the bandwidth of the type II radio burst and the peak proton flux, and proposed that the bandwidth represents the strength of the shock wave. In our examples, the time-averaged bandwidth for Event~1 was $>$~1000~kHz, wider than that for Event~2 ($<$~500~kHz), suggesting that the shock wave in Event~2 was weaker.

\subsection{Properties of the Active Regions}
\label{AR}

%%%%%%%%%%%%%%%%%%%%%%%%%%%%%%%%%%%%%%%%
\begin{figure*}
\gridline{\fig{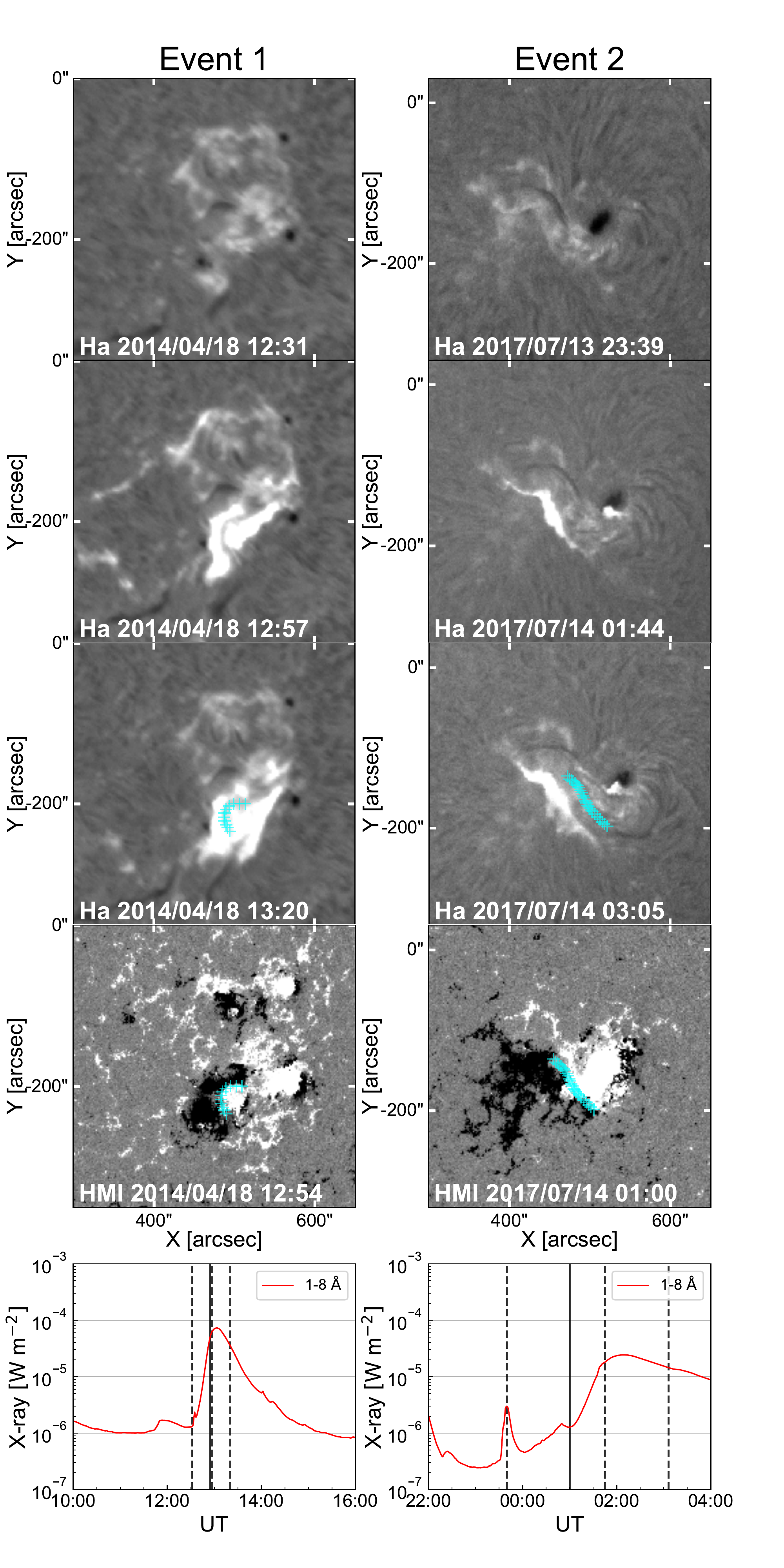}{0.5\textwidth}{}}
\caption{The observations of the solar surface at each event. The top three panels are H$\mathrm{\alpha}$ ground-based observations, and the bottom is HMI line-of-sight magnetogram.
The bottom panel is a 1\,--\,8$\mathrm{\mathring{A}}$ light curve of SXR, and the black dashed and solid lines correspond to the observation times of H$\mathrm{\alpha}$ and the magnetic field, respectively.
The cyan cross markers indicate the location of the top of the post-flare loop with reference to the flare ribbon and polarity inversion line.
}
\label{fig:halpha}
\end{figure*}
%%%%%%%%%%%%%%%%%%%%%%%%%%%%%%%%%%%%%%%%

We discuss how different proton release times may be traced back to the properties of the active regions that produced the CMEs.
Figure \ref{fig:halpha} shows H$\alpha$ images and magnetograms of the active regions that produced the two events (AR~12036 and AR~12665). For each of the regions, the top three rows display H$\alpha$ images taken at three times (before and around the flare peak and during the decay phase) that are indicated by black dashed lines on the GOES 1\,--\,8~\AA\ light curves in the bottom panels. The fourth row gives a line-of-sight magnetogram from the Helioseismic and Magnetic Imager \citep[HMI;][]{scherrer_2012_sp} on board SDO,  taken in the early phase of the flare (see the black solid line in the bottom row). The H$\alpha$ data come from the Global Oscillation Network Group \citep[GONG\footnote{\url{https://gong.nso.edu}};][]{harvey_1996_sci} for Event~1 and the Solar Dynamics Doppler Imager \citep[SDDI;][]{ichimoto_2017_sp} installed on the Solar Magnetic Activity Research Telescope \citep[SMART;][]{ueno_2004_aspc} at Hida Observatory.  

We readily note from the magnetograms that the region for Event~1 is more magnetically complex than the one for Event~2, which is dominated by essentially a simple bipolar topology.  This difference is also noted in the pre-flare H$\alpha$ images (the top row of Figure~\ref{fig:halpha}). The complex magnetic field configuration of the region for Event~1 may be reflected also in the complex evolution of the flare ribbons in H$\alpha$ images as shown in the second and third rows of Figure~\ref{fig:halpha}. However, the apparent difference of the complexity of the two regions may not be reflected in basic magnetic parameters from the Space-Weather HMI Active Region Patches \citep[SHARP;][]{bobra_2014_sp} over several (e.g., 6~hour, 12~hour, 24~hour) intervals preceding the flare onsets.  None of them seem to distinguish the two regions in a significant way. 

Flare ribbons contain additional information of flares. Concerning our examples, the initial distance of the flare ribbons for Event~1 is shorter ($\sim$20~Mm) than that for Event~2 ($\sim$50~Mm). This is consistent with the result that those with widely separated ribbons in the beginning tend to be of long duration \citep{toriumi_2017_apj}.
Accordingly, the flare loops are longer in the region for Event~2 than in the region for Event~1, which may translate to a higher initial reconnection point in Event~2. The magnetic field strength near the reconnection point is, therefore, expected to be weaker in Event~2, suggesting that it could not drive the faster ejection near the solar surface.

Another information we can get from the area of flare ribbons is the reconnection flux, which may be related to the photospheric magnetic flux traversed by the flare ribbons \citep[e.g.,][]{forbes_1984_sp, kazachenko_2017_apj}. Analyzing the flare ribbons in AIA 1600~\AA\ images, 
\cite{kazachenko_2017_apj} created a database of the reconnected flux of 3137 $\gtrsim$C1 flares up to April 2016.  The reconnection flux of the flare for Event~1 is 9.44$\times$10$^{21}$Mx, according to the database. A new calculation shows that for Event~2 to be 2.92$\times$10$^{21}$Mx (M. Kazachenko, 2021, private communication). 
However, the reconnection rate normalized by the duration of the flare is 3.21$\times$10$^{18}$~Mx~s$^{-1}$ in Event~1 and 3.53$\times$10$^{17}$~Mx~s$^{-1}$ in Event~2, which is smaller by one order of magnitude.

%%%%%%%%%%%%%%%%%%%%%%%%%%%%%%%%%%%%%%%%
%%%%%%%%%%%%%%%%%%%%%%%%%%%%%%%%%%%%%%%%
\begin{figure*}
\gridline{\fig{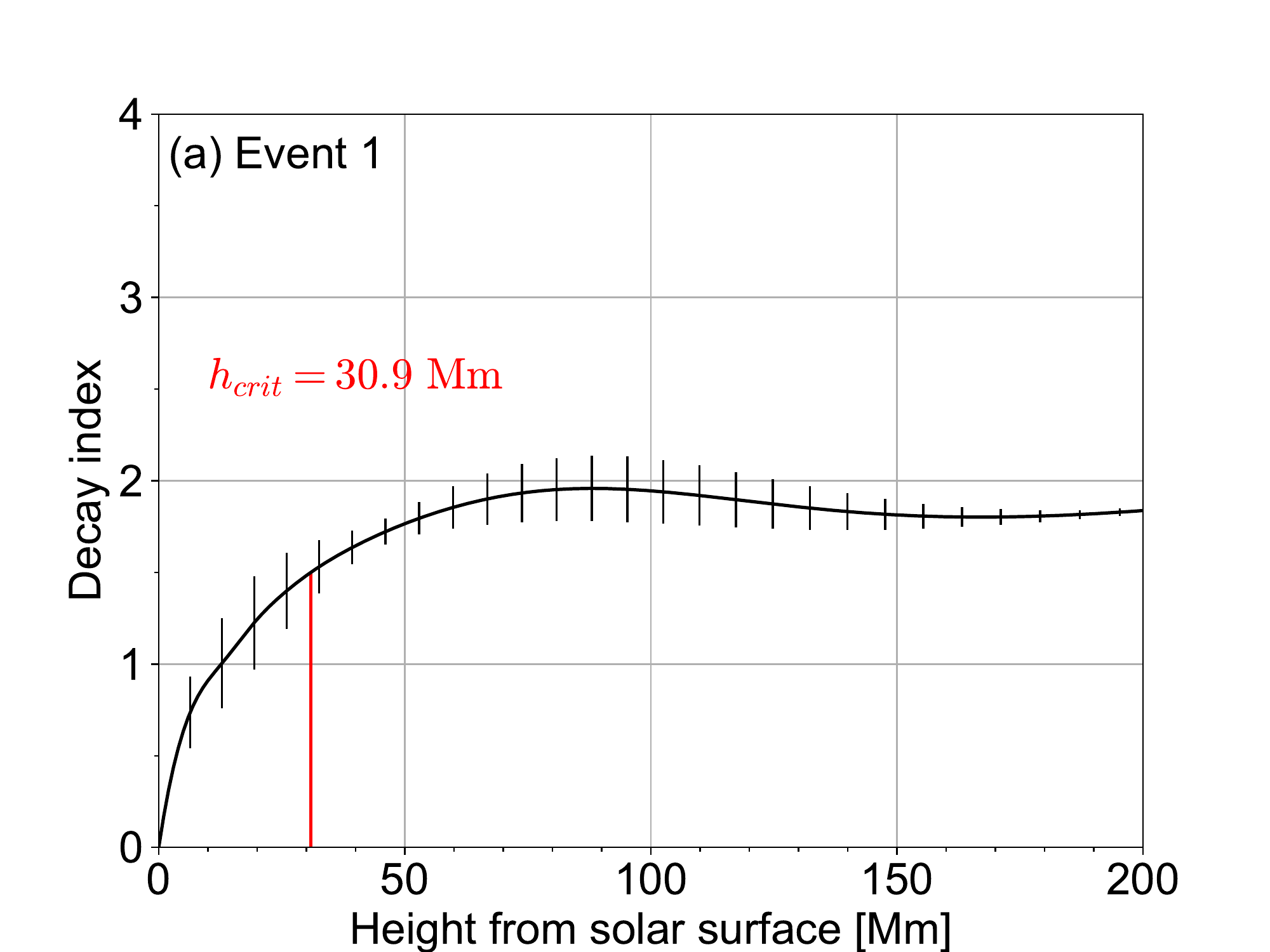}{0.49\textwidth}{}
          \fig{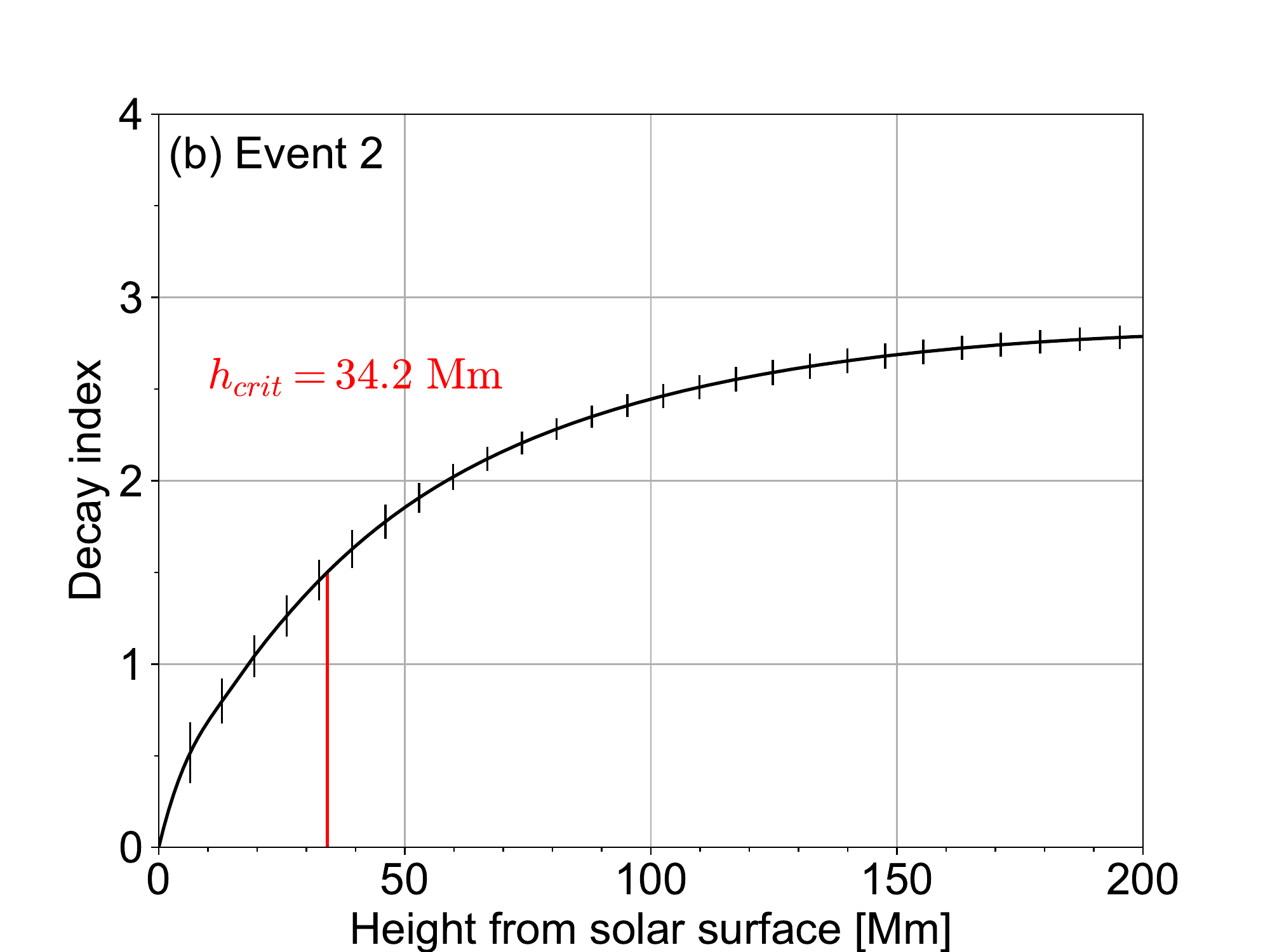}{0.49\textwidth}{}}
\caption{The decay index vs height for the regions responsible for (a) Event~1 and (b)  Event~2.
The red line corresponds to the critical height at which the decay index becomes 1.5. The error bars show the standard deviation at each height.}
\label{fig:di}
\end{figure*}
%%%%%%%%%%%%%%%%%%%%%%%%%%%%%%%%%%%%%%%%
%%%%%%%%%%%%%%%%%%%%%%%%%%%%%%%%%%%%%%%%

Lastly, to address the possible difference of the overlying magnetic structure in the regions responsible for Event~1 and Event~2, we calculate the decay index, $n = -d \mathrm{ln} B/d \mathrm{ln} h$, which shows how quickly the magnetic field weakens with height over the polarity inversion lines that align with the tops of the post-flare loops (cyan marks in the third and fourth rows of Figure~\ref{fig:halpha}).
The PFSS model\footnote{\url{https://github.com/dstansby/pfsspy/} developed by \citet{stansby_2020_joss}} is used to calculate the coronal magnetic field. The bottom boundary is the downsized (720$\times$360 pixels) standard HMI Carrington synoptic map, embedded with the original HMI magnetogram of an area of (200$\arcsec$)$^2$ around the core of the active region, which is taken just before the flare.
The decay index gives a criterion for torus instability to trigger an eruption 
\citep{kliem_2006_phrvl}. In Figure~\ref{fig:di}, the decay index over the height up to 200~Mm is shown for the two regions. We find almost no difference in the so-called critical height ($h_{crit}$) at which $n_{crit}$=1.5. 

However, the decay index above 50~Mm tends to be larger in the region for Event~2, meaning the magnetic field decreases more quickly with height. Indeed, $|B_{r}|$ at the source surface of 2.5~R$_{\odot}$, which is the upper boundary of the calculation domain, is smaller in Event~2 region than that in Event~1 region.
We note that even the value in Event~1
is smaller than calculated with any of the magnetic field models that are used in Section~\ref{sec:shock}. A similar finding was reported for example by \cite{rouillard_2016_apj}, consistent with the smaller open flux at 1~AU as predicted with photospheric magnetograms than actually observed \citep[e.g.,][]{linker_2017_apj}.

%%%%%%%%%%%%%%%%%%%%%%%%%%%%%%%%%%%%%%%%%%%%%%%%%%%%%%%%%%%%%%%%%%%%%%%%%%%%%%%%
% Discussion and Analysis
%%%%%%%%%%%%%%%%%%%%%%%%%%%%%%%%%%%%%%%%%%%%%%%%%%%%%%%%%%%%%%%%%%%%%%%%%%%%%%%%

\section{Summary and Conclusions}
\label{sec:summary}

In this work, we choose two events that had widely different TO, the time of the SEP onset (found in GOES $>$10~MeV proton data) from the onset of the CME, although they occurred at similar longitudes and were associated with CMEs that had similar speeds \citep{kihara_2020_apj}. After reviewing SOHO/ERNE data, we decided to use these data with much lower background and better energy discrimination. The revised TO, or particle (proton) release time (as obtained with a VDA) from the onset of the type II radio burst, shows a 
smaller difference between the two events. However, the difference of 20\,--\,25 minutes is still significant, unaccounted for by different path lengths (1.34~AU vs 1.44~AU, see Figure~\ref{fig:vda}) from the VDA. 

In order to understand the longer delay of the proton release time in Event~2, we focus on how the shock wave grows close to the Sun as characterized by the Alfv\'{e}n Mach number $M_{A}$ of the shock waves, by more closely examining the height-time profiles of the CMEs than single fits over the entire FOV covered by LASCO data. Despite strong model dependency of the Alfv\'{e}n speed $v_{A}$ especially on the magnetic field, $M_{A}$ keeps rising and reaches certain thresholds around the proton release times.  It has been shown with more sophisticated tools \citep[e.g.,][]{rouillard_2016_apj, kouloumvakos_2019_apj} that protons are released when $M_{A}$ reaches a critical value.  We note that slow acceleration of the CME over long time while the soft X-ray flux was on the rise is a key to the delayed acceleration/release of protons in Event~2. 

Another possibility is that the shock wave driven by the CME in Event~2 was intrinsically weak, as suggested by small bandwidths of the type II radio burst \citep[cf.,][]{iwai_2020_apj}, not being capable of accelerating protons on its own, but that interaction with the previous CME may have been instrumental in the production of energetic protons \citep[e.g.,][]{gopalswamy_2002_apj, li_2005_icrc, li_2012_ssrv}.  The timing of the possible CME-CME interaction is consistent with the proton release as far as LASCO imagery is concerned, but radio signatures come an hour later. This may not be a problem until we better understand at what timing during CME-CME interactions we expect to observe the radio signatures.

In either case, we try to find different properties of the active regions that hosted the CMEs. The region for Event~2 had much simpler magnetic field configurations, consistent with the way flare ribbons developed. The eruption involved a larger volume, producing a flare that lasted for more than a day. Even though it is not straightforward to extract the possible differences of active regions in the forms of the routinely calculated magnetic field properties with HMI data \citep[SHARP;][]{bobra_2014_sp}, reconnection flux from flare ribbons \citep{kazachenko_2017_apj}, or the decay index \citep{kliem_2006_phrvl}, the overall simple magnetic configurations allowed slow but steady acceleration of the CME in Event~2. They should also be conducive to the long-duration flare.
Although the energy release was not intense at first, the injection of magnetic energy lasted for a long time. Eventually a shock wave strong enough to generate SEPs was formed, or a CME fast enough to catch up with the former one was formed.

In this paper, we try to show that the acceleration of CMEs and the growth of the Alfv\'{e}n Mach number below $\sim$10 R$_\odot$ may play a significant role in the onset of SEPs. This needs to be verified in a larger sample of events. To make such an attempt meaningful, it would be vital to characterize the distribution of magnetic field and density in this height range beyond the utilization of simple models, which may be helped by MHD simulations. Transport processes such as cross field diffusion, which may be at work even in the height range of interest, could affect the onset behaviors of SEP events. To evaluate the effect of such processes, we would need more detailed modeling of heliospheric magnetic field. In addition, we speculate on the basis of our findings that there may be a connection between the complexity of active regions and the timescales of flares and CMEs, which may more directly affect the temporal characteristics of SEPs. This presents one interesting possibility for a comprehensive explanation of these solar active phenomena.

We close with a cautionary note that GOES EPS data may not be suitable for scientific analyses of onset times in particular for events like our Event~2, where protons increase slowly from a low level. The smaller difference in the proton release time as found using ERNE data may explain only marginal differences in active region properties.  We suggest that the past and ongoing results based on GOES EPS data should be calibrated with other data.

\begin{acknowledgments}
This study is based on the discussion at the Coordinated Data Analysis Workshops held in August 2018 and 2019 held under the auspice of the Project for Solar-Terrestrial Environment Prediction \citep[PSTEP;][]{kusano_2021_eps}. We thank the reviewer for their helpful comments on the manuscript.
This work was supported by JSPS KAKENHI grant No. JP22J11442 (K.K.) and JP21H01131 (A.A.) and also by the joint research project of the Unit of Synergetic Studies for Space, Kyoto University and BroadBand Tower, Inc. (BBT). The work of N.V.N. was supported by NASA grants 80NSSC18K1126 and 80NSSC20K0287.

\end{acknowledgments}

\bibliography{bibliography}{}
\bibliographystyle{aasjournal}

\end{document}